\documentclass[10pt,a4paper]{article}

\usepackage{PRIMEarxiv}
\usepackage[utf8]{inputenc} 
\usepackage[T1]{fontenc}    
\usepackage{hyperref}       
\usepackage{url}            
\usepackage{booktabs}       
\usepackage{amsfonts}       
\usepackage{nicefrac}       
\usepackage{microtype}      
\usepackage{lipsum}
\usepackage{fancyhdr}       
\usepackage{graphicx}       
\graphicspath{{media/}}     

\usepackage{amsmath}
\usepackage[numbers,sort&compress]{natbib}
\usepackage{xurl}   
\usepackage{rotating} 
\usepackage{listings} 
\usepackage{xcolor}   

\hypersetup{
    colorlinks=true,
    linkcolor=blue,
    filecolor=magenta,      
    urlcolor=cyan,
    citecolor=red,
}

\definecolor{codegreen}{rgb}{0,0.6,0}
\definecolor{codegray}{rgb}{0.5,0.5,0.5}
\definecolor{codepurple}{rgb}{0.58,0,0.82}
\definecolor{backcolour}{rgb}{0.95,0.95,0.92}
\lstdefinestyle{mystyle}{
    backgroundcolor=\color{backcolour},   
    commentstyle=\color{codegreen},
    keywordstyle=\color{magenta},
    numberstyle=\tiny\color{codegray},
    stringstyle=\color{codepurple},
    basicstyle=\ttfamily\footnotesize,
    breakatwhitespace=false,         
    breaklines=true,                 
    captionpos=b,                    
    keepspaces=true,                 
    numbers=left,                    
    numbersep=5pt,                  
    showspaces=false,                
    showstringspaces=false,
    showtabs=false,                  
    tabsize=2
}
\title{Sentinel: SOTA model to protect against prompt injections

}

\author{
  Dror Ivry \\
  Qualifire \\
  Tel Aviv, IL\\
  \texttt{dror@qualifire.ai} \\
   \And
  Oran Nahum \\
  Qualifire \\
  Tel Aviv, IL\\
  \texttt{oran.nahum@qualifire.ai}
}

\begin{document}
\maketitle

\begin{abstract}
Large Language Models (LLMs) are increasingly powerful but remain vulnerable to prompt injection attacks, where malicious inputs cause the model to deviate from its intended instructions. This paper introduces Sentinel, a novel detection model, \texttt{qualifire/prompt-injection-sentinel}, based on the \texttt{answerdotai/ModernBERT-large} architecture. By leveraging ModernBERT's advanced features and fine-tuning on an extensive and diverse dataset comprising a few open-source and private collections, Sentinel achieves state-of-the-art performance. This dataset amalgamates varied attack types, from role-playing and instruction hijacking to attempts to generate biased content, alongside a broad spectrum of benign instructions, with private datasets specifically targeting nuanced error correction and real-world misclassifications. On a comprehensive, unseen internal test set, Sentinel demonstrates an average accuracy of 0.987 and an F1-score of 0.980. Furthermore, when evaluated on public benchmarks, it consistently outperforms strong baselines like \texttt{protectai/deberta-v3-base-prompt-injection-v2}. This work details Sentinel's architecture, its meticulous dataset curation, its training methodology, and a thorough evaluation, highlighting its superior detection capabilities.
\end{abstract}

\keywords{Prompt Injection \and Large Language Models \and LLM Security \and ModernBERT \and Jailbreak}

\newpage

\section{Introduction}
Large Language Models (LLMs) have revolutionized numerous domains, showcasing remarkable capabilities in understanding, generation, and agentic applications. However, a critical vulnerability inherent in many LLMs is their difficulty in distinguishing between trusted system instructions and untrusted user-provided data. This vulnerability is exploited by prompt injection attacks, where malicious inputs manipulate the model into ignoring its original instructions and executing unintended, often harmful, actions \cite{greshake2023notwhat, perez2022ignore}. The fundamental issue lies in the LLM's instruction-following ability being co-opted.

Existing detection methods for prompt injection face several limitations. Trained detectors, while sometimes effective, can exhibit biases towards their training data, leading to poor generalization on unseen attack vectors. For instance, the \texttt{protectai/deberta-v3-base-prompt-injection} model, while demonstrating impressive results on some benchmarks, showed significantly lower performance when evaluated on a more diverse dataset, suggesting a potential training data bias. The dynamic nature of attacks, often described as a "cat-and-mouse game", means that defenses trained on known attack signatures can quickly become outdated.

This paper introduces Sentinel, a robust detection model officially named \texttt{qualifire/prompt-injection-sentinel} \cite{qualifireSentinelModelCard}. It is built by Qualifire upon the \texttt{answerdotai/ModernBERT-large} architecture \cite{warner2024modernbert, answeraiModernBERTCard}. Sentinel is specifically trained to classify prompts as either benign or indicative of a jailbreak/injection attempt. By leveraging a meticulously aggregated and diverse dataset, Sentinel demonstrates superior performance. The primary contributions of this work are: (1) The development and comprehensive evaluation of Sentinel. (2) The curation of a diverse training and testing dataset. (3) A comparative analysis showcasing Sentinel's significant improvements over established baselines on both a comprehensive internal test set and public benchmarks.

\section{Related Work}

\subsection{Prompt Injection Attacks}
Prompt injection attacks aim to subvert an LLM's intended purpose by embedding malicious instructions within the input prompt \cite{greshake2023notwhat, perez2022ignore}. These attacks can range from simple commands like "Ignore previous instructions" to more sophisticated techniques involving role-playing, obfuscation, character escaping, or context ignoring \cite{alkaswan2025isyourpromptsafe}. Recent research highlights various attack patterns, including direct injection, indirect prompt injection, and jailbreaks designed to bypass safety alignments \cite{zhang2025oet, alkaswan2025isyourpromptsafe, anon2025evolving}. The continuous evolution of attack methods, including template-based, generative, and optimization-driven approaches \cite{anon2025evolving}, necessitates robust detection mechanisms.

\subsection{Prompt Injection Defenses and Detection}
A primary approach to mitigation is the development of trained detector models. These are typically binary classifiers fine-tuned on datasets containing examples of both benign and malicious prompts. Notable examples include models based on architectures like DeBERTa \cite{he2021deberta}, such as \texttt{protectai/deberta-v3-base-prompt-injection} and its subsequent versions \cite{protectaiDebertaV1, protectaiDebertaV2}. Other defense strategies include adversarial training and input preprocessing \cite{zhang2025oet}. The effectiveness of trained detectors heavily depends on the quality and diversity of their training data. Recent studies emphasize the importance of dynamic benchmarks for evaluating defenses, as even strong cutting edge models can exhibit vulnerabilities \cite{zhang2025oet, anon2025evolving}.

\section{The Sentinel Detection Model}

\subsection{Model Architecture: Leveraging ModernBERT}
Sentinel utilizes \texttt{answerdotai/ModernBERT-large} as its base. ModernBERT is a modernized bidirectional encoder-only Transformer model pretrained on 2 trillion tokens of English and code data, with a native context length of up to 8,192 tokens \cite{warner2024modernbert}. The "large" variant consists of 28 layers and 395 million parameters. Key architectural features include:
\begin{itemize}
    \item \textbf{Rotary Positional Embeddings (RoPE):} \cite{su2021roformer} For superior long-context support and relative position encoding.
    \item \textbf{Local-Global Alternating Attention:} For efficient processing of long input sequences.
    \item \textbf{Unpadding and Flash Attention:} \cite{dao2022flashattention} For optimized inference speed and memory efficiency.
\end{itemize}

\subsection{Dataset Curation and Preparation}
A key component of Sentinel's development was the creation of an extensive and diverse dataset for training and evaluation.

\subsubsection{Open-Source Datasets}
The following open-source datasets were incorporated:
\begin{itemize}
    \item \texttt{Salad-Data} \cite{opensafetylabSaladData}: Filtered for 'O5: Malicious Use' category, this dataset is known for including creative and complex jailbreak attempts.
    \item \texttt{alespalla/chatbot-instruction-prompts} \cite{alespallaChatbotPrompts}: A source of benign prompts (7,000 samples used).
    \item \texttt{microsoft/orca-agentinstruct-1M-v1} \cite{microsoftOrcaAgentInstruct}: Benign prompts (7,000 samples extracted from the "content" field).
    \item \texttt{verazuo/jailbreak-llms} \cite{SCBSZ24}: A collection of jailbreak and benign prompts from an in-the-wild repository.
    \item \texttt{lmsys/toxic-chat} \cite{lmsysToxicChat}: Used for jailbreak prompts where the "jailbreaking" column indicated an attack.
    \item \texttt{VMware/open-instruct} \cite{vmwareOpenInstruct}: A source of benign prompts (7,000 samples used).
    \item \texttt{reshabhs/SPML-Chatbot-Prompt-Injection} \cite{reshabhsSPML}: Contains ~16,000 samples focusing on scenario-based attacks.
\end{itemize}

\subsubsection{Private Datasets}
To address specific challenges, Qualifire developed a private dataset:
\begin{itemize}
    \item \texttt{qualifire-synthetics}: Contains 1,400 synthesized using LLMs.
\end{itemize}

\subsubsection{Final Dataset Composition and Splitting}
After consolidating all sources, the dataset was structured to have approximately 70\% benign and 30\% jailbreak prompts. The entire dataset was then split into a 90\% training set and a 10\% test set, ensuring no overlap between them.

\subsection{Training}
Sentinel was developed by fine-tuning the \texttt{answerdotai/ModernBERT-large} model on the 90\% training split of the curated dataset. The task was formulated as a binary classification problem.

\section{Experimental Setup}

\subsection{Test Dataset}
Evaluation was conducted on two fronts: (1) the 10\% held-out internal test set, comprising a diverse mix of prompts from all source datasets, and (2) several public, standardized prompt injection benchmarks.

\subsection{Baseline Model}
The primary baseline for comparison is \texttt{protectai/deberta-v3-base-prompt-injection-v2} \cite{protectaiDebertaV2}.

\subsection{Evaluation Metrics}
For the internal test set, we used Accuracy (AvgAcc), Recall, Precision, and F1-score. For public benchmarks, we report the Binary F1 Score as is standard.

\section{Results and Analysis}

\subsection{Performance on Internal Test Set}
On our comprehensive internal test set, Sentinel demonstrated significantly superior performance compared to the baseline. The results are summarized in Table~\ref{tab:internal_performance}.

\begin{table}[htbp]
  \centering
  \caption{Performance on the Internal Held-Out Test Set}
  \label{tab:internal_performance}
  \resizebox{\columnwidth}{!}{%
  \begin{tabular}{@{}lccccc@{}}
    \toprule
    Model & AvgAcc & Recall & F1 & Precision & Params \\
    \midrule
    \textbf{qualifire/prompt-injection-sentinel} \cite{qualifireSentinelModelCard} & \textbf{0.987} & \textbf{0.991} & \textbf{0.980} & \textbf{0.986} & 0.395B \\
    protectai/deberta-v3-base-prompt-injection-v2 \cite{protectaiDebertaV2} & 0.848 & 0.905 & 0.728 & 0.820 & 0.185B \\
    \bottomrule
  \end{tabular}%
  }
\end{table}

Sentinel achieved an average accuracy of 0.987, surpassing the baseline by 13.9 percentage points. The F1-score of 0.980 is substantially higher than the baseline's 0.728. The superior performance is attributed to both the advanced ModernBERT architecture and the extensive and diverse training dataset.

\subsection{Benchmark Performance}
To further validate Sentinel's robustness and generalization, we evaluated it on four challenging public prompt injection benchmarks. As shown in Table~\ref{tab:benchmark_performance_resized}, the Sentinel model consistently and significantly outperforms the strong DeBERTa-v3 baseline across all datasets.

\begin{table*}[htbp]
  \centering
  \caption{Benchmark Performance (Binary F1 Score)}
  \label{tab:benchmark_performance_resized}
  \resizebox{\textwidth}{!}{%
    \begin{tabular}{@{}lccccc@{}}
      \toprule
      Model & \shortstack{\texttt{allenai/wildjailbreak}\\\cite{allenaiWildjailbreak}} & \shortstack{\texttt{jackhhao/jailbreak}\\\texttt{-classification}\\\cite{jackhhaoJailbreakClassification}} & \shortstack{\texttt{deepset/prompt}\\\texttt{-injections}\\\cite{deepsetPromptInjections}} & \shortstack{\texttt{qualifire/Qualifire-prompt}\\\texttt{-injection-benchmark}\\\cite{qualifireSentinelBenchmarkDataset}} & \textbf{Avg} \\
      \midrule
      \texttt{qualifire/prompt-injection-sentinel} \cite{qualifireSentinelModelCard} & \textbf{0.935} & \textbf{0.985} & \textbf{0.857} & \textbf{0.976} & \textbf{0.938} \\
      \texttt{protectai/deberta-v3-base-prompt-injection-v2} \cite{protectaiDebertaV2} & 0.733 & 0.915 & 0.536 & 0.652 & 0.709 \\
      \bottomrule
    \end{tabular}%
  }
\end{table*}

The results on these public benchmarks confirm the findings from our internal test set. Sentinel's average F1 score of 0.938 is nearly 23 points higher than the baseline's 0.709. This consistent out performance provides strong evidence of the model's superior generalization capabilities.

\subsection{Latency and  Hardware}
This benchmark was performed using an L4 GPU. Due to the extremely small size of the model the latency was quite surprising at an avg latency of just \textasciitilde0.02 seconds per inference time we achieved incredible real-time evaluation capabilities on a fairly modest hardware.

\section{Discussion and Limitations}
The results indicate that Sentinel sets a new standard for open-source prompt injection detection. The improvement over the baseline highlights the benefits of using advanced base models and investing in extensive dataset curation.

Despite its strong performance, Sentinel has limitations:
\begin{enumerate}
    \item \textbf{Susceptibility to Novel Attacks:} As a trained model, its knowledge is based on its training data. Highly novel attack vectors could potentially evade detection \cite{zhang2025oet}.
    \item \textbf{Dataset Reproducibility:} The inclusion of private datasets means that exact replication of the training environment is contingent on access to these proprietary assets.

\end{enumerate}

\subsection{Error analysis}

To better understand Sentinel’s limitations, we conducted a manual review of a random sample of misclassifications from the internal test set. Surprisingly, the errors did not fall into distinct or recurring categories. Instead, we didn't observed any specific identifiers or characteristocs for the errors observed.
False positives (benign prompts incorrectly classified as injections) - typically included edge cases involving unusual formatting, assertive or security-related phrasing, or ambiguous intent. False negatives (missed jailbreaks) - tended to involve subtle adversarial phrasing that did not strongly resemble known attack patterns.

\section{Conclusion and Future Work}
Sentinel has demonstrated state-of-the-art performance in prompt injection detection. This success underscores the importance of advanced model architectures and comprehensive training data.

Future work will focus on:
\begin{itemize}
    \item \textbf{Continuous Dataset Evolution:} Regularly updating the training dataset with new jailbreak techniques.
    \item \textbf{Model Optimization:} Exploring techniques like knowledge distillation and quantization to create smaller, faster versions of Sentinel.
    \item \textbf{Hybrid Defense Approaches:} Investigating the integration of Sentinel with other defense mechanisms, such as input sanitization or runtime monitoring \cite{anon2025evolving}.
\end{itemize}

\appendix

\section{How to Get Started with the Model}

This section provides a simple code snippet to get started with the \texttt{qualifire/prompt-injection-sentinel} model using the Hugging Face \texttt{transformers} library. Ensure you have the library installed (\texttt{pip install transformers torch}).

\subsection{Code Example}
\begin{lstlisting}[language=Python, caption=Python code to run the Sentinel model.]
from transformers import pipeline, AutoTokenizer, AutoModelForSequenceClassification

model_id = 'qualifire/prompt-injection-sentinel'

# Load the tokenizer and model from Hugging Face Hub
tokenizer = AutoTokenizer.from_pretrained(model_id)
model = AutoModelForSequenceClassification.from_pretrained(model_id)

# Create a text-classification pipeline
pipe = pipeline("text-classification", model=model, tokenizer=tokenizer)

# Test with a benign prompt
result = pipe("hi how are you?")

print(result)
\end{lstlisting}

\subsection{Example Output}
The code above will produce the following output for a benign prompt, indicating a high confidence score for the 'benign' label.
\begin{lstlisting}[language=bash, backgroundcolor=\color{white}]
[{'label': 'benign', 'score': 1.0}]
\end{lstlisting}

\bibliographystyle{plainnat}
\bibliography{references}

\end{document}